\documentclass[nobm]{epl}
\newcommand{\be}{\begin{equation}}
\newcommand{\ee}{\end{equation}}
\newcommand{\ba}{\begin{eqnarray}}
\newcommand{\ea}{\end{eqnarray}}

\title{Reactive dynamics on fractal sets: anomalous
fluctuations and memory effects}
\shorttitle{Reactive dynamics on fractal sets}

\author{E. Abad\inst{1}\thanks{E-mail:\email{
eabad@ulb.ac.be}} \and A. Provata\inst{2} \and G. Nicolis\inst{1}}
\institute{
\inst{1} Centre for Nonlinear Phenomena and Complex Systems, 
Universit\'e Libre de Bruxelles CP 231, 1050 Bruxelles, Belgium.\\
\inst{2} Institute of Physical Chemistry, National Research Center 
''Demokritos'', 15310 Athens, Greece.
}

\begin{document}

\maketitle

\begin{abstract}
We study the effect of fractal initial conditions in closed
reactive systems in the cases of both 
 mobile and immobile reactants. For the reaction $A+A\rightarrow A$, 
in the absence of diffusion, the mean number of particles $A$
is shown to decay exponentially to a steady state
which depends on the details of the initial conditions.
The nature of this dependence is demonstrated both analytically and 
numerically. In contrast, when
diffusion is incorporated, it is shown that the mean number of 
particles $\langle N(t)\rangle$ 
decays asymptotically as $t^{-d_f/2}$,
the memory of the initial conditions being now carried by the dynamical 
power law exponent. The latter is fully determined by  
the fractal dimension $d_f$ of the initial conditions.
\end{abstract}

\section{Introduction}

In classical approaches to reactive processes mean-field (MF) theories
have been useful to predict many non-trivial dynamical
and steady state properties such as multistability, periodic concentration
oscillations, chaotic motion etc. As long as fluctuations in 
concentration and occupation number space remain small, MF approaches provide 
a suitable description of the process. This is the case in many practical 
situations, where inhomogeneous fluctuations can be efficiently removed from 
the system by continuous external stirring of the reactants or 
sufficiently fast internal diffusion \cite{nicpr}. 

On the other hand, considerable interest has been devoted to reactive 
processes taking place on low-dimensional supports in recent years. In such 
systems, external stirring may prove difficult, and the 
internal diffusional mixing may not be sufficiently fast to compensate 
the effect of the restricted support geometry \cite{slink}. 
Correlated spatial fluctuations initially present in the system or induced 
by the interplay between the chemistry and the spatial characteristics of 
the support may then give rise to anomalous temporal behaviour 
\cite{prak} and even deviations from the MF steady state \cite{prov1}.
In particular, the outcome of the reaction may be strongly influenced by 
the dimensionality of the support. 

In recent years, the important role of dimensionality was recognized to
be an issue common to many statistical processes as well 
\cite{prov1,privb}. Fractal supports, which can be 
produced to display a continuum of dimensionalities, have been used
widely in experimental \cite{kop1, kop2} and 
theoretical studies \cite{tretya} to illustrate the new effects
arising in the above context. A second factor of importance in the
interplay between local dynamics and transport processes is the form of
the initial conditions \cite{linden}. Depending on whether they are 
homogeneous, short-ranged or
correlated on a long range they may favour quick mixing by diffusion, 
delays in such a mixing resulting in segregation or, in the limit of
immobile reactants, nonergodic behaviour altogether. The objective of
the present work is to analyze the evolution of a low dimensional system
subject to initial conditions corresponding to a fractal spatial 
distribution of reactants in the lattice. In particular, 
the role of the fractal dimension $d_f$ of the initial 
distribution will be assessed.   

The specific system we shall analyze here is the 
reaction $A+A \rightarrow A$ \cite{spou, avrabursch}. The interest in
the dynamics of this reaction on fractal supports first arised in 
connection with exciton fusion experiments on fractal percolation clusters
\cite{kop1, kop2}. While previous work has treated the case 
of random homogeneous initial conditions \cite{baras} 
and the general inhomogeneous case \cite{abad1}, we focus here on 
the specific case of an inhomogeneous fractal distribution.  
We consider the cases where a) the reactants $A$ are immobile on a 
fractal subset of a 1D lattice and b) the reactants $A$ are initially 
placed on the fractal subset, but for $t>0$ they can diffuse and react
throughout the entire 1D lattice. We show, analytically and numerically, that
for immobile reactants the steady state reactant concentration does not 
depend only on $d_f$ but also on the details of the fractal subset. In 
contrast, for diffusing reactants we show that the memory of the initial 
fractal distribution is carried by the anomalous dynamical exponent which 
is fully determined by $d_f$. 

\section{Dynamics on fractals: immobile reactants}

To investigate the effect of fractal initial conditions on reactive systems
with immobile reactants we consider the reaction
$A+A \rightarrow A$ on two different
Cantor-like sets $C_1$ and $C_2$. Set $C_1$ is
obtained by iteration of the segment 1110,
while set $C_2$  is formed by  iteration of
the segment 1101.
After the $n$-th iteration, the lattice size $L$ defined for both
sets contains $4^n$ sites. In both sets, the number of ones and zeros  
is $3^n$ and $4^n-3^n$ respectively. We consider the "1" sites as the 
active sites (sites where reactions can take place) and the "0" sites as 
inactive, or empty sites. Both sets are deterministic fractals 
with $d_f=\log(3)/\log(4)$. For convenience, we shall
number the lattice sites from $1$ to $L$, say from left to right,
regardless of whether they are "0" (inactive) or "1" (active) sites.

Next, we fill the "1" sites of set $C_1$ or set $C_2$ with particles $A$ and 
let the reactions proceed according to the following Monte Carlo (MC) 
algorithm: at each time step $\Delta t$, a lattice site $i$ is randomly 
chosen. If $i$
and a randomly chosen neighbour $i\pm 1$ are occupied, the particle $A$ at site
$i$ is removed from the lattice with probability $k_R$. In our model, the site
choice is unbiased, {\it i.e.} sites $i+1$ and $i-1$ are chosen with equal
probability. On the other hand, $k_R$ is the conditional probability
of reaction at each time step given that two neighbouring sites are occupied
and plays thus the role of a local reaction rate. We choose fixed
boundary conditions by introducing two additional sites $0$ and
$L+1$ at the boundaries and specifying them as "0" sites.
This particular choice of boundary conditions is selected for
convenience and does not a play any important role in the large $L$
limit. Finally, we set $\Delta t=\frac{1}{L}$
\footnote{ This choice implies 
that all lattice sites are scanned once on average after 
one time unit.} 

We are interested in the time evolution of
the mean particle number $\langle N(t) \rangle$ and the associated 
mean concentration $\theta(t):=
\langle N(t) \rangle/L$.  
Clearly, with the above choice of initial conditions, 
$\langle N(0) \rangle$ diverges while 
$\theta(0)$ vanishes in the limit $L\to\infty$ as 
$L^{(\log(3)/\log(4))-1}=L^{d_f-1}$. 
In order to obtain a well posed problem, we 
shall therefore consider the case of a finite 
system, as it is the case in experimental situations
involving, for instance, mesoscopic scale devices (micelles, single
crystallographic faces of a solid catalyst, etc.). Our objective 
will be to see whether some generic trends will nevertheless show up 
for long times and/or large sizes $L$. We expect
that the final number of particles $\langle N(\infty) \rangle$
will strongly depend on the specific form of the initial condition
due to the lack of mixing, since particle islands evolve independently from
each other. Thus, the spatial correlations present in the initial
distribution will propagate in time.

The function $\langle N(t) \rangle$ can be obtained
heuristically by means of the following observation \cite{cohen}: 
a string of $k$ consecutive sites ($k$-tuplet) can be destroyed either
by internal reaction events between particles inside the tuplet
or by reaction between the particles at each edge site of the
tuplet and particles sitting at occupied nearest neighbour sites
outside the $k-$tuplet. The latter events require the existence
of a $k+1$-tuplet. The dynamics of the mean number of
$k-$tuplets $M^{(n)}_k(\tau)$ (averaged over an ensemble of
identical lattices) in a fractal set of size
$L=4^n$ is given by the following hierarchy of equations \cite{baras,abad1}:
\be
\label{tuplev}
\frac{d}{d\tau} M^{(n)}_k(\tau)=-(k-1)M^{(n)}_k(\tau)
-M^{(n)}_{k+1}(\tau), \qquad k=1\cdots k_{max},
\ee
where $k_{max}$ is the size of the largest $k$-tuplet and
$M^{(n)}_k(\tau)\equiv 0$ for $k>k_{max}$. The reaction
rate $k_R$ has been absorbed in the adimensional time variable
$\tau=k_R\,t$. The general solution of eqs. (\ref{tuplev}) 
depends strongly on the details of the fractal set. It reads
\be
\label{gentupsol}
 M^{(n)}_k(\tau)=e^{-(k-1)\tau}\sum_{s=0}^{k_{max}-k}
\frac{(e^{-\tau}-1)^s}{s!}M^{(n)}_{k+s}(0).
\ee
For sets $C_1$ and $C_2$, $k_{max}=3$. The mean number of particles (singlets)
is then
\be
\label{singdyn}
\langle N(\tau)\rangle =M^{(n)}_1(\tau)=\sum_{s=0}^2
\frac{(-1)^s}{s!}M^{(n)}_{1+s}(0)+(M^{(n)}_2(0)-M^{(n)}_3(0))\,e^{-\tau}
+\frac{M^{(n)}_3(0)}{2}\,e^{-2\tau}.
\ee
In particular, the initial tuplet distribution for set $C_1$ is given
by the number of particle islands ($=3^{n-1}$) times the number of
$k$-tuplets contained by each island.  In this case, there 
are three singlets, two doublets and one triplet in each island, 
{\it i.e.}
$M^{(n)}_1(0)=3\cdot 3^{n-1}=3^n,\, M^{(n)}_2(0)=2\cdot 3^{n-1}
\mbox{ and } M^{(n)}_3(0)=1\cdot 3^{n-1}$. Asymptotically, we have
\be
\label{asnum}
\langle N(\infty)\rangle =\sum_{s=0}^2
\frac{(-1)^s}{s!}M^{(n)}_{1+s}(0)=\frac{3^n}{2}=\frac{
\langle N(0)\rangle}{2}
=\frac{3}{2}\cdot 3^{n-1}.
\ee
This gives a survival factor $\eta(n):= 
\langle N(\infty)\rangle/\langle N(0)\rangle=\frac{1}{2}$, {\it i.e.} 
the number of particles drops to half the
initial value regardless of the lattice size. As emphasized 
by eq. (\ref{asnum}), {\it each} island
yields asymptotically a mean number of particles equal to $\frac{3}{2}$.
In contrast, a simple-minded combinatorial counting giving each
final state of the island (ASA),(SAS),(SSA),(ASS) the same statistical
weight yields the wrong factor $\frac{5}{4}$. This reflects the 
nonergodicity of the system, implying that the number 
of statistical paths leading to each steady state is different.

For set $C_2$, the initial distribution is slightly more complex and
contains a variety of island sizes. We have $M^{(n)}_1(0)=3^n,\quad 
M^{(n)}_2(0)=\frac{1}{2}(3^n-1) \mbox{ and } 
M^{(n)}_3(0)=\frac{1}{2}(3^{n-1}-1)$.
Using again eq. (\ref{singdyn}), we obtain $\langle N(\infty) \rangle=
\frac{7}{12}\,3^n+\frac{1}{4}$ and $\eta(n)=\frac{7}{12}+
\frac{3^{-n}}{4}$. In the large $L$ limit, 
$\eta(n)\to\frac{7}{12}\approx 0.583$. The size distribution of
the islands plays a crucial role to determine the number of
surviving particles. Particles in smaller islands have a higher
survival expectancy. Therefore, more particles survive in set
$C_2$ than in set $C_1$.

For set $C_2$, the combinatorial argument is again based on the
size distribution of the islands, which is connected with
the tuplet distribution through the equation 
$I^{(n)}_k(\tau)=M^{(n)}_k(\tau)-2M^{(n)}_{k+1}(\tau)+
M^{(n)}_{k+2}(\tau)$ \cite{abad1}.
We then have $I^{(n)}_1(0)=I^{(n)}_2(0)=\frac{1}{2}(3^{n-1}+1) 
\mbox{ and } I^{(n)}_3(0)=\frac{1}{2}(3^{n-1}-1)$.
According to the combinatorial counting, islands of size $1$ and
$2$ will yield one particle asymptotically, whereas island of size $3$
yield $\frac{5}{4}$ particles on average. This is again
wrong since, as we know, each three-particle island 
is reduced to $\frac{3}{2}$ particles on average. 

The time evolution of $\langle N(\tau)\rangle$ is easily computed
for both fractal sets by substituting the corresponding expressions
for the $k$-tuplet distributions into eq. (\ref{singdyn}). In the long 
time limit,
the dominant term describing the decay to the steady state
is proportional to $e^{-\tau}$ (recall that $\tau=k_R\,t$). Thus, the 
information on the initial
distribution is contained in the coefficient of the dominant term
rather than in the relaxation time $k_R^{-1}$ given by the exponent of 
the dominant term. In particular, this means that fractals with different 
$d_f$ may relax at the same speed into the steady state. We therefore conclude
that in the immobile reactant case neither the dynamics nor the 
steady state are suitably characterized by $d_f$.

To confirm our results, we have performed MC simulations over 
$2 \cdot 10^3$ statistical runs on a lattice with $L=4^5=1024$ sites. For both 
fractal sets, the asymptotic concentrations $\theta(\infty)$ 
and the dynamics dictated by eq. (\ref{singdyn}) agree very well
with the simulations (fig. \ref{fig1}). 

Let us compare the previous results for inhomogeneous initial conditions
with the case of a lattice containing only "1'' sites. A lattice of
length $L=4^n$ can be regarded as an iteration of, say, the segment
$1111$. The initial $k$-tuplet distribution is given by 
$M^{(n)}_k(0)=4^n-k+1$ with $k=1,..,4^n$. In the large $L$ limit,
this yields 
$\langle N(\infty)\rangle=\sum_{s=0}^{4^n-1} (4^n-s)/s!\approx e^{-1}(4^n-1)$
and $\eta(n)\approx e^{-1}(1-4^{-n})$, {\it i.e.}, $\eta(n)\to
e^{-1}\approx 0.367$ as $L^{-1}$. As expected, the homogeneous system is
characterized by a lower survival factor than any Cantor-like
set for all values of $L$.

A comment on the new features brought in by the fractal initial conditions
analyzed above is now in order. The initial inhomogeneities 
imposed by sets $C_1$ or $C_2$ decouple the dynamics of different
parts of the system and decompose it into smaller, homogeneous subsystems.
In principle, they have the same effect as the reaction-induced 
inhomogeneities, {\it i.e.} lowering the number of active sites. 
However, the interest of considering inhomogeneous initial conditions
lies in the fact that, while the initial spatial correlations
range over the whole system size, chemically-induced correlations 
are short-ranged; the latter may only develop between sites initially 
belonging to the same island. Due to the absence of diffusion or any
other randomizing mechanism, a detailed memory of the initial 
spatial structure is carried by $\theta(\tau)$ for all times. 

As found above, the system's memory in this case is not sufficiently 
characterized by $d_f$. The mean coordination number $z$ of the fractal 
subset, defined as the spatial average of the number of active neighbour 
sites, is not adequate either for the description the dynamics or 
the steady state. Indeed, the fractal set obtained by the iteration of 
the segment $1110011110011000$ has the same value of $d_f$ and 
$z=\frac{4}{3}$ as set $C_1$, but its dynamics and steady state are
not the same due to the different $k$-tuplet distribution. 


An alternative analytical description for the reaction
on linear sets is provided by the theory of Markov chains.
We have seen that the reactive dynamics of, say, set $C_1$, can be fully
determined by knowledge of the evolution of a single three-particle
island. An island can never evolve into the empty state (SSS). There
are 7 possible states \footnote{The number
of relevant states can be decreased by symmetry considerations, but
we shall keep all seven states for the sake of clarity.} of such an island 
with at least one particle, namely (AAA), (SAA),(ASA),(AAS),(SSA),(SAS) 
and (ASS). Let us denote them by 1,2,3,4,5,6 and 7 respectively. The 
state vector of the system at time $t$ is $\vec{P}(t)=(P_1(t),\cdots, 
P_7(t))^T$, where $P_n(t)$ is the probability 
that the island be in the state $n$ at time $t$.
$\vec{P}(t)$ satisfies the stochastic evolution equation  
\be
\label{matdyn}
\vec{P}(t+\Delta t)={\bf W}^T \vec{P}(t),
\ee
where the elements $w_{i,j}$ of
the transition probability matrix {\bf W} are computed from the 
MC algorithm for the reaction by counting the number of paths 
leading from one state to another: 
\be
{\bf W}=\left( \begin{array}{*{7}c} 1/3 & 1/6 & 1/3 & 1/6 & 0 & 0 & 0 \\
                  0 & 2/3 & 0 & 0 & 1/6 & 1/6 & 0 \\
                  0 & 0 & 1 & 0 & 0 & 0 & 0 \\
                  0 & 0 & 0 & 2/3 & 0 & 1/6 & 1/6 \\
                  0 & 0 & 0 & 0 & 1 & 0 & 0 \\
                  0 & 0 & 0 & 0 & 0 & 1 & 0 \\
                  0 & 0 & 0 & 0 & 0 & 0 & 1 \end{array} \right)
\ee
(we have set $k_R=1$ for simplicity). This matrix possesses four 
absorbing states and is therefore nonergodic \cite{feller}.
The steady states $\vec{\Pi}$ are the eigenvectors of the
matrix ${\bf W}^T$ corresponding to the degenerate eigenvalue $\lambda_1=1$.
The algebraic multiplicity of $\lambda_1$ is $4$, equal to the dimension of
the subspace spanned by the four eigenstates $\vec{\Pi}_i$ with 
the components $\Pi_{1,j}=\delta^{kr}_{3,j},
\Pi_{2,j}=\delta^{kr}_{5,j},\Pi_{3,j}=\delta^{kr}_{6,j}$
and $\Pi_{4,j}=\delta^{kr}_{7,j}$, where $j=1,\ldots,7$.
The general form of the steady state $\vec{\Pi}$ can be expressed as a
superposition of the states $\vec{\Pi}_i$, {\it i.e.} $
\vec{\Pi}=\sum_i c_i \,\vec{\Pi}_i \mbox{ with } \sum_i c_i=1
$.
The coefficients $c_i$  depend on the initial conditions and
can be calculated numerically by iteration of the evolution equation
(\ref{matdyn}). Thus, if one starts with a full island (state 1),
the asymptotic mean number of particles on the island will be
$\langle N(\infty)\rangle_{isl.}
=2\cdot \frac{1}{2}+ 1\cdot \frac{1}{8}+1\cdot
\frac{1}{4}+1\cdot \frac{1}{8}=\frac{3}{2}$, 
{\it i.e.} we recover the result obtained previously for the Cantor set $C_1$.
$\langle N(\infty)\rangle$ can be computed similarly for the 
set $C_2$ and for any other Cantor-like sets. 

The other eigenvalues of ${\bf W}^T$ describe the decay of
$\langle N(t)\rangle$ to the steady state, 
{\it i.e.} $\langle N(t)\rangle-\langle N(\infty)\rangle
=\sum_{i=2}^3 u_i\,e^{-\lambda_i\,t}$ with time-independent coefficients
$u_{2,3}$. They are $\lambda_2=\frac{2}{3}$ (twice)
and $\lambda_3=\frac{1}{3}$. This time dependence is corroborated 
by eq. (\ref{singdyn}), 
except that the arguments in the exponential functions differ from
our result by a factor of $\frac{1}{3}$. However, this artefact is a direct 
consequence of the choice for the time unit: in the model for the 
$k$-tuplets and in the MC simulations, it was the time 
needed to scan the whole lattice, while here it has been implicitly assumed
to be the time required to 
update a {\it single} site in a three-site lattice.   

\section{Dynamics on fractals: diffusing reactants}

Reactive events in the above diffusionless systems can be viewed as particle
jumps into already occupied sites. One can also allow for
additional diffusion events, {\it i.e.} particles can jump into the "0" sites
of the fractal sets with a probability rate $k_D$. The "0" sites
can then no longer be considered as inactive sites but rather as initially 
empty sites. A particle initially placed on a "1" site of the fractal
set can now diffuse into a "0" (empty) neighbour site. The latter 
becomes then a "1" site (occupied), while the original site becomes 
"0" (empty). On long time scales, this gives classical
diffusion. A description in terms of islands is no longer suitable, 
since they can now interact with each other by means of diffusing 
particles at the boundaries of each island.  

To study the long-time dynamics, we have performed a series of MC 
simulations. Initially, particles are put on the "1" sites
of the lattice and then they start to diffuse freely over the entire lattice
and react with each other. In this case we take reflecting boundary
conditions, {\it i.e.} when particles arrive at sites $0$ or $L+1$, they bounce
off and go back to sites $1$ and $L$, respectively. Again, the time scale
is set by $L^{-1}$. The reaction rate $k_R$, taken to be equal to 
the diffusion rate $k_D$, can be again absorbed in the time scale by 
setting $\tau=k_R\,t$. 

\begin{figure}[htbp]
\twofigures[scale=0.35]{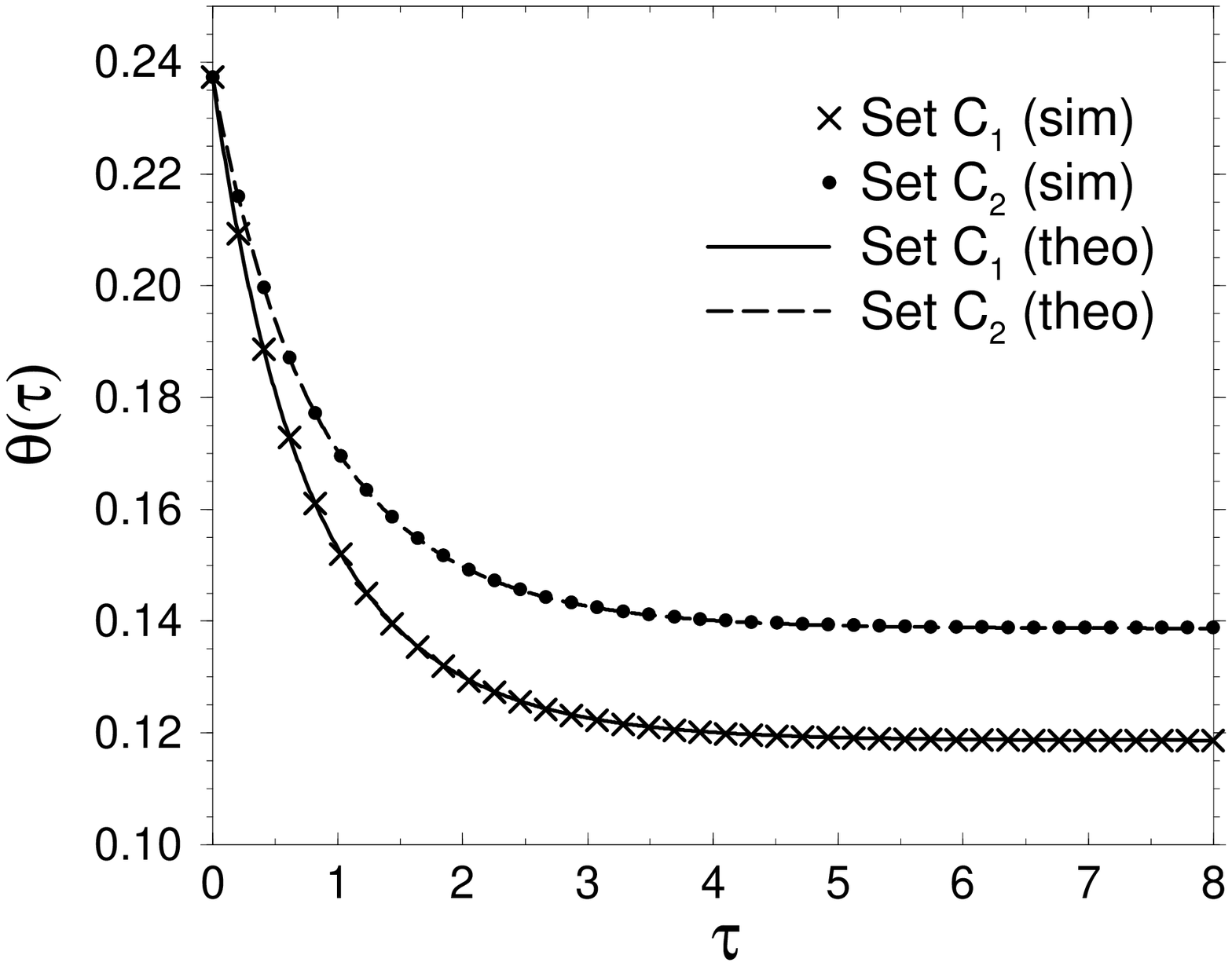}{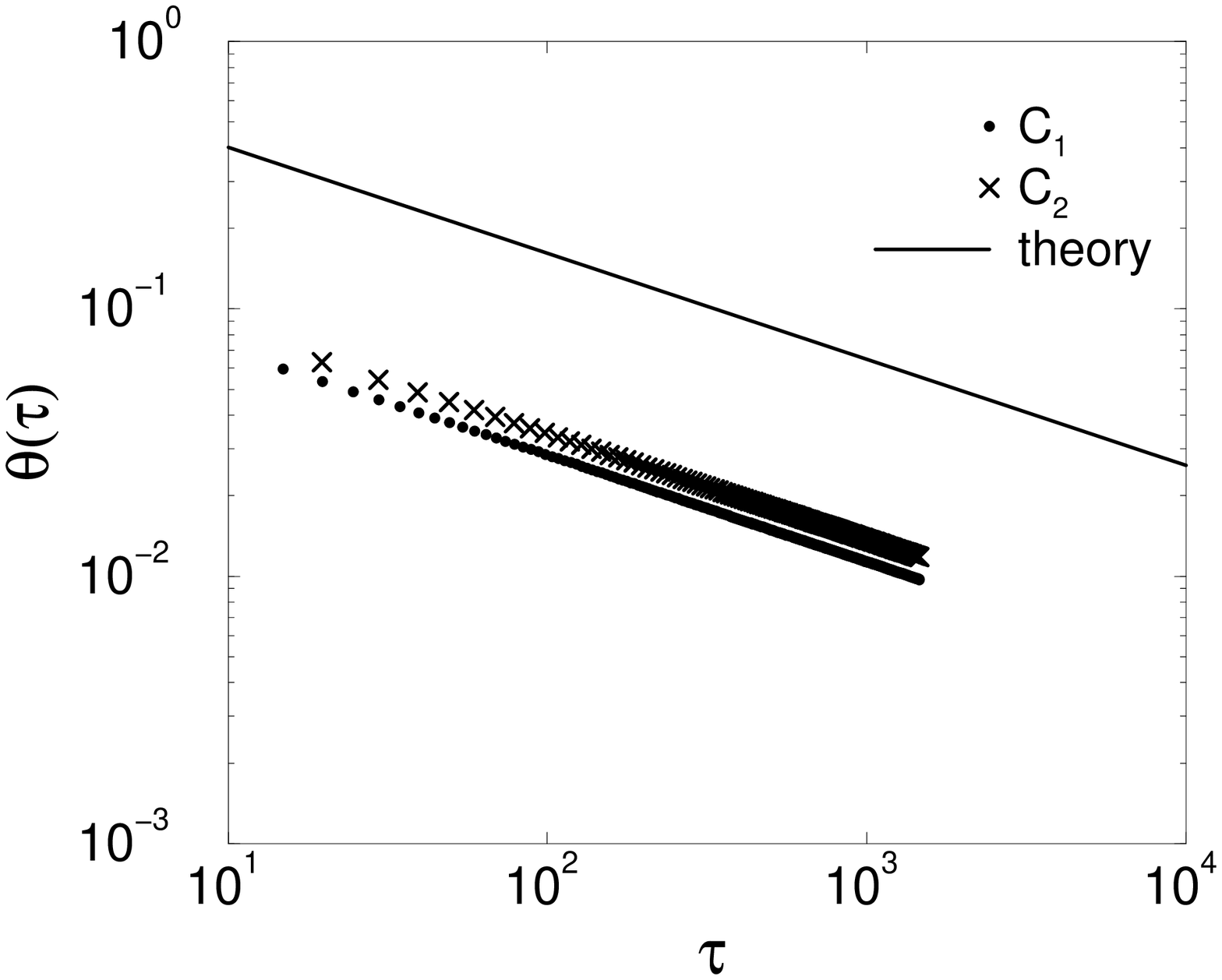}
\caption{Time evolution of the particle concentration 
in the absence of diffusion. The dots and crosses represent the MC results,
while the continuous and dashed curves have been computed 
from eq. (\ref{singdyn}).}
\label{fig1}
\caption{Time evolution of the particle concentration
in the presence of diffusion.
The slope of the continuous line is equal to $-d_f/2$, {\it i.e.} the 
theoretical value of the dynamical exponent (see text).}
\label{fig2}
\end{figure}

For comparison with the diffusionless case,
we have used as initial conditions the fractal sets $C_1$
and $C_2$ and have performed the same number of runs ($2\cdot 10^3$) 
for $L=4^5$. The
time evolution of the concentration $\theta(\tau)$ has been 
monitored for sets $C_1$ and $C_2$. Figure \ref{fig2} displays a
double-logarithmic plot of numerical MC results for 
$\theta(\tau)$ as a function of $\tau$. The dynamical 
exponent is given by the slopes of the point 
curves, which are represented by dots for $C_1$ and crosses for $C_2$; for 
both sets, the mean slope is very close to the value 
$-d_f/2$ represented by the continuous straight line. 
In contrast to the immobile reactant case, 
$d_f$ appears to be the natural parameter associated to the long time 
decay of the concentration. 

The above analysis suggests a power law behaviour 
$\theta(t)=\alpha\,t^{-d_f/2}$ governing the long time dynamics. This law
exhibits two types of memory effects: a memory of the dimensionality 
$d_f$ of the initial particle distribution entering via the exponent;
and a more detailed memory of the initial distribution, entering via 
the amplitude factor $\alpha$. It is only in the regime of extremely
long times $t \gg L^2$ that the decay curves for sets $C_1$ and $C_2$
fall into each other. This limit is to some extent
trivial from the standpoint of many-particle dynamics, since 
only one particle remains on the lattice.
In the long time regime, detailed information about 
lacunae in the initial particle distribution will also 
be kept before reaching the steady state in higher order quantities 
like the distribution function for the interparticle distance. 

These results agree with recent theoretical predictions. 
When $k_D=k_R$, a closed analytic description in terms of 
empty $k$-tuplets (also termed intervals) is possible \cite{avrabursch,prov2}. 
This method has been used to show that, for fractal 
initial conditions with $d_f<2$, $\langle N(t)\rangle$ behaves as 
$t^{-d_f/2}$ at long times in the large $L$ limit \cite{prov2,alem}; for
$d_f=2$, logarithmic corrections are necessary, while MF theory applies
for $d_f>2$. In our case $d_f=\log(3)/\log(4)\approx 0.792$, so the
exponent ''remembers'' the initial particle distribution. 

\section{Conclusions}

In the present work the reaction process $A+A\rightarrow A$ with fractal 
initial conditions has been studied both for immobile and diffusing reactants.
In the diffusionless case, it was shown that the fractal dimension $d_f$ 
does not suffice to characterize the dynamics and the steady state. The 
number of surviving particles at the steady state depends on the details
of the initial distribution. In the presence of diffusion, 
multistationarity is suppressed and the steady state becomes universal 
and MF-like. However, 
a long tail characteristic of anomalous dynamics subsists, implying that
the decay is governed by a power law $\langle N(t)\rangle \propto t^{-d}$
rather than an exponential. In contrast to the immobile 
reactant case, the memory of the initial condition is 
carried by the characteristic dynamical
exponent $d$ rather than by the steady state, where $d$ is fully determined
by $d_f$. As expected, the memory of the initial condition is less detailed
than in the immobile reactant case, due to the randomizing effect of
diffusion.  

The above results may be relevant for a series of experimental situations 
involving systems other than particle aggregates. A few examples are 
heterogeneous catalysis, evaporation-deposition systems, porous media, 
percolation clusters and ferromagnetic systems. 

Our work can be generalized in many different ways. One can {\it e.g.} consider
the case of random rather than deterministic fractal initial conditions. 
Other possibility is studying more complex reactive schemes like the
reversible case $A+A\rightleftharpoons A$ \cite{prak}, 
whose long-time dynamics still remains to be characterized in 
detail in the limit of immobile reactants.  

\acknowledgements

We thank K. Karamanos, Dr. A. Shabunin, Prof. V. Astakhov, G. A. 
Tsekouras, H. L. Frisch, F. Vikas and F. Baras for helpful discussions. 
This work was supported, in part, by the NATO 
Collaborative Linkage Grant No PST.CLG.977654 and the Interuniversity 
Attraction Poles program of the Belgian Federal Government.

\newpage

\end{document}